\documentclass[11pt]{article}
\usepackage[a4paper,margin=1in]{geometry}
\usepackage[T1]{fontenc}
\usepackage{amsmath,amssymb,amsthm}
\usepackage[hidelinks]{hyperref}

\newcommand{\ALG}{\mathrm{ALG}}
\newcommand{\OPT}{\mathrm{OPT}}
\newcommand{\E}{\mathbb{E}}
\newcommand{\tn}[1]{\left\lVert#1\right\rVert}
\newtheorem{theorem}{Theorem}[section]
\newtheorem{lemma}[theorem]{Lemma}
\newtheorem{proposition}[theorem]{Proposition}

\title{Constant Competitive Ratios for \\Stochastic Online Metric Matching}
\author{Chenglin Fan\thanks{Seoul National University.} \and Mingyu Kang\footnotemark[1]}
\date{}
\begin{document}
\maketitle
\begin{abstract}
In stochastic online metric matching, $n$ server locations are
known in advance, and $n$ requests are drawn independently from
a common distribution. Each request must be matched to an unused
server upon arrival. The goal is to minimize the total matching
cost. We prove that Fair-Bias, the algorithm of Gupta, Guruganesh,
Peng, and Wajc~\cite{GGPW}, is $700$-competitive on any metric
space when each request arrives at the location of a uniformly
random server. The previous best bound was
$O((\log\log\log n)^2)$. When the request distribution is unknown
and no prior samples are available, our result implies that
Harada's algorithm~\cite{Harada} is $4201$-competitive on finite
metric spaces. By projecting requests to their nearest servers,
we obtain a competitive ratio of $8403$ on arbitrary metric
spaces. Our proof samples servers independently and integrates
the expected transportation cost over the sampling probability.
We group nearby servers and use the transportation cost at a
lower sampling rate to control the contribution of large local
sample counts. We also extend a bound on the expected cost of
a single match from tree metrics to general metrics, up to a
constant factor.
\end{abstract}

\section{Introduction}

Online metric matching is a classical problem in online algorithms.
We are given $n$ servers in a metric space $(X,d)$, and $n$ requests
arrive one at a time. The server locations are known in advance.
Each server has a label $i\in[n]=\{1,\ldots,n\}$ and a location
$s_i\in X$, and several labels may have the same location.
Each request must be irrevocably matched to an unused server upon
arrival, at a cost equal to the distance between them. The goal is
to minimize the total matching cost.

In the stochastic model, the requests are drawn independently from
a common Borel probability distribution $D$ on $X$. In the known
i.i.d.\ model, the algorithm is given $D$ in advance. In the unknown
i.i.d.\ model, the algorithm neither knows $D$ nor receives any
prior samples. Let $\ALG$ denote the total cost of the online
algorithm and $\OPT$ the minimum cost of a perfect matching for
the realized requests. An algorithm is $\alpha$-competitive if
$\E[\ALG]\leq \alpha\,\E[\OPT]$ for every instance in the model under
consideration with finite $\E[\OPT]$. The expectations are over
the requests and the random choices of the algorithm.

\subsection{Previous Work}

For adversarial arrivals, Kalyanasundaram and Pruhs~\cite{KP93}
and Khuller et al.~\cite{KMV94} independently established the optimal
deterministic competitive ratio of $2n-1$. Meyerson et al.~\cite{MNP06}
gave a randomized algorithm with competitive ratio $O(\log^3 n)$,
which was improved to $O(\log^2 n)$ by Bansal et al.~\cite{BBGN14}.
Gupta and Lewi~\cite{GL12} obtained a ratio of $O(\log n)$ on metrics
of bounded doubling dimension. In the random order model, an adversary
chooses the requests, but they arrive in a uniformly random order.
Raghvendra~\cite{Rag16} established a tight competitive ratio of
$\Theta(\log n)$ in this model. Thus, randomizing the arrival order
alone does not suffice for a constant competitive ratio on general
metrics.

For known i.i.d.\ arrivals, Gupta, Guruganesh, Peng, and Wajc~\cite{GGPW}
introduced Fair-Bias and obtained a competitive ratio of $9$ on
tree metrics and $O((\log\log\log n)^2)$ on general metrics. Their
analysis on trees uses the mass imbalance across each tree edge.
For general metrics, they combine an implicit tree embedding argument
with bounds on the occupancy counts of server locations. Different
bounds are used depending on the number of servers that remain free.
Although this gives a much smaller upper bound than in the random
order model, it does not establish a constant competitive ratio.
In Section~5 of their full paper, they asked whether Fair-Bias, or
any other online algorithm, could achieve a constant competitive
ratio on general metrics for known or unknown i.i.d.\ arrivals.

Constant competitive ratios have also been obtained under more
restrictive assumptions on the input. Balkanski et al.~\cite{BFP23}
proved such a bound for Greedy when both servers and requests are
sampled independently and uniformly from the unit interval. Yang
and Yu~\cite{YYGreedy26} extended the constant bound to fixed
Euclidean dimensions at least three and showed that Greedy has
competitive ratio $\Theta(\sqrt{\log n})$ in dimension two under
uniform sampling. Saberi et al.~\cite{SYY24} and Yang and Yu~\cite{YY26}
reduced the known distribution model with arbitrary server placements
to a model with randomly sampled servers. Their results include
constant competitive ratios on the Euclidean unit cube in fixed
dimension at least three for smooth distributions. Li et
al.~\cite{LVY26} obtained constant competitive ratios for independent
requests from possibly different smooth distributions in fixed
Euclidean dimensions other than two, given one sample from each
distribution. These results exploit geometric or distributional
assumptions that need not hold in general metric spaces.

Recently, Harada~\cite{Harada} obtained a competitive ratio of
$O((\log\log\log n)^2)$ on general metrics and a constant competitive
ratio on tree metrics in the unknown i.i.d.\ model. His guarantees
match those of Gupta et al.~\cite{GGPW} up to constant factors. His
reduction also shows that a constant competitive ratio for Fair-Bias
would imply a constant competitive ratio without knowledge of the
request distribution. Thus, a constant bound for Fair-Bias on general
metrics would resolve the open question in both i.i.d.\ models.

\subsection{Our Results}

We resolve the open question of Gupta et al.~\cite{GGPW}
for both known and unknown i.i.d.\ arrivals, without any geometric
restriction on the metric or smoothness assumption on the request
distribution.

\begin{theorem}\label{thm:unknown}
There is a constant-competitive randomized algorithm for online
metric matching with unknown i.i.d.\ arrivals on every metric space.
\end{theorem}

The algorithm is due to Harada~\cite{Harada}.
We obtain a competitive ratio of $4201$ on finite metric spaces
and $8403$ on arbitrary metric spaces.
The latter bound follows by projecting each request to its
nearest server among all original servers, with ties broken
in a fixed order~\cite{Harada,MNP06}.
Since every matching costs at least $\OPT$, the optimal
competitive ratio in the unknown i.i.d.\ model is $\Theta(1)$.
Together with the logarithmic lower bound for random order
arrivals~\cite{Rag16}, this establishes an asymptotically tight
separation between the two models.

Both the reductions of Gupta et al.~\cite{GGPW} and
Harada~\cite{Harada} lead to the same special case.
Let $\delta_x$ denote the unit point mass at $x$.
Define the \emph{server distribution}
\[
    \nu=\frac{1}{n}\sum_{i=1}^{n}\delta_{s_i}.
\]
A request drawn from $\nu$ arrives at the location of a uniformly
chosen server label.
Our main technical result is the following theorem.

\begin{theorem}\label{thm:canonical}
Fair-Bias is $700$-competitive on every metric space when requests
are drawn independently from the server distribution $\nu$.
\end{theorem}

Suppose Fair-Bias is $\alpha$-competitive under the server
distribution. Harada's reduction~\cite[Theorem~6.2]{Harada}
then gives a competitive ratio of $1+6\alpha$ for unknown
i.i.d.\ arrivals on finite metric spaces. Setting $\alpha=700$
gives a competitive ratio of $4201$.
Similarly, the reduction of Gupta et al.~\cite[Lemma~2.1]{GGPW}
gives a competitive ratio of $1401$ for known i.i.d.\ arrivals
on finite metric spaces.

\subsection{Technical Overview}

The main difficulty is to bound the expected transport cost of a
uniformly random subset of servers in terms of the expected offline
optimum. The tree analysis of Gupta et al.~\cite{GGPW} benefits from
an exact decomposition of transport costs along edges. Using a tree
embedding to transfer this analysis to general metrics can incur a
logarithmic distortion~\cite{FRT04}. Their occupancy bounds avoid
this loss in some ranges of the number of free servers, but the
combined estimates still give a bound that grows with $n$. We
instead work directly in the original metric and express the total
expected cost as an integral over a sampling probability. This
lets us use independence between sample counts in disjoint groups
and compare costs at different sampling probabilities.

\paragraph{From random subsets to independent sampling.}
Fair-Bias computes an optimal transport plan from the server
distribution to the uniform distribution on the free server
labels. When requests are drawn from $\nu$, each free label
is chosen with the same probability. Consequently, for every
$1\leq k\leq n$, the set $F_k$ of remaining server labels is
a uniformly random subset of $[n]$ of size $k$~\cite[Lemma~2.2]{GGPW}.
Let $W_1(\mu,\eta)$ denote the minimum transport cost between
finitely supported probability distributions $\mu$ and $\eta$. Define
\[
    \nu_{F_k}=\frac{1}{k}\sum_{i\in F_k}\delta_{s_i},
    \qquad
    g_k=\E[W_1(\nu_{F_k},\nu)].
\]
Then $g_k$ is the expected cost of the next match when $k$
servers remain free, and the total expected cost of Fair-Bias
is $\sum_{k=1}^n g_k$. For a subset of fixed size, the sample
counts in disjoint groups are dependent. We instead sample
each server label independently with probability $t\in[0,1]$.
Let $J_i(t)$ be the indicator that label $i$ is sampled and
put $v_i=\delta_{s_i}-\nu$. For a finitely supported signed measure $z$ of total
mass zero, let $\tn{z}$ be the minimum cost of transporting its
positive part to its negative part. In particular, each $v_i$
has total mass zero and $\sum_i v_i=0$. Define
\[
    H(t)=\E\!\left[\tn{\sum_{i=1}^{n}J_i(t)v_i}\right].
\]
Conditioned on sampling exactly $k\geq 1$ labels, the sample
is a uniformly random subset of size $k$ and its expected
transport cost is $kg_k$. The cost is zero for an empty sample.
Integrating over the sampling probability gives
\[
    \E[\ALG]=\sum_{k=1}^{n}g_k
    =\int_0^1 H(t)\frac{dt}{t},
\]
where $\ALG$ is the cost of Fair-Bias under $\nu$.
It therefore suffices to bound $H(t)$ by a quantity that
decreases polynomially as $t$ tends to zero. Let
$\varepsilon_1,\ldots,\varepsilon_n$ be independent random
signs, each uniform on $\{-1,1\}$, and define
\[
    R=\E\!\left[\tn{\sum_{i=1}^{n}\varepsilon_i v_i}\right].
\]
Let $\OPT_\nu$ be the offline optimum for $n$ independent
requests from $\nu$. A comparison between sampling with
and without replacement gives $R\leq 2\E[\OPT_\nu]$.

\paragraph{Grouping nearby servers.}
First assume that $n\geq 2$ and the server locations are distinct.
For an integer $1\leq q\leq n$, let $r_i(q)$
be the distance from server $i$ to its $q$th nearest site,
counting itself, and let $L(q)=\sum_{i=1}^n r_i(q)$.
By constructing Lipschitz functions with separated supports,
we show that $L(q)\leq 8qR$ for $2\leq q\leq\lfloor n/2\rfloor$.
For such $q$, we partition the sites into
groups of size between $q$ and $2q$. Spreading one unit of
mass from every site uniformly within its group costs at
most $4L(q)$ in total. Each site is sampled with probability $t$, so performing these moves
only at sampled sites costs at most $4tL(q)$ in expectation.
After this redistribution, the remaining transport cost
depends only on the sample count $K_G$ in each group $G$.
The counts are independent, and $\E[K_G]=t|G|\leq 4$ when
$tq\leq 2$. We truncate each count at $24$ and use
symmetrization and a comparison with sums of independent
signs to bound the contribution of the centered truncated
counts by $24\sqrt{3/q}\,R$. This leaves the contribution
of counts above $24$.

\paragraph{Controlling large sample counts.}
We express the remaining part of each centered group count
as a sum of centered indicators. For each integer
$\ell\geq25$, the corresponding indicator records whether
$K_G\geq\ell$ and is centered by subtracting its expectation.
Set $c_\ell=(4e/\ell)^\ell$. Take a fresh sample by including each site independently with probability
$c_\ell t$. We show that replacing the centered indicator for $G$
by twice the centered count in $G$ from this sample can only increase
the expectation of a convex function. The transport norm is
convex in each coefficient, and the group counts are
independent, so we can apply this comparison one group at
a time. The contribution of level $\ell$ is then at most
$2H(c_\ell t)$. Combining the three parts gives
\[
    H(t)\leq
    32tqR+24\sqrt{\frac{3}{q}}R
    +2\sum_{\ell\geq25}H(c_\ell t)
\]
for $0<t\leq1$, integers $2\leq q\leq\lfloor n/2\rfloor$,
and $tq\leq2$. Balancing the first two terms suggests
choosing $q=\max\{2,\lceil t^{-2/3}\rceil\}$.
If $q>n/2$, a direct bound suffices.
Otherwise, the first two terms in the recurrence are of order $t^{1/3}R$. Since
$2\sum_{\ell\geq25}c_\ell^{1/3}<1/16$, the remaining terms can
be absorbed into a uniform bound on $H(t)/(t^{1/3}R)$. This gives
$H(t)\leq114t^{1/3}R$ for every $0<t\leq1$. Using the integral identity and
$R\leq2\E[\OPT_\nu]$ gives
\[
    \E[\ALG]\leq342R\leq684\,\\E[\OPT_\nu].
\]
A perturbation of the distances extends the result to
coincident server locations, giving Theorem~\ref{thm:canonical}.

\paragraph{The cost of a single match.}
In Appendix~\ref{sec:sharp}, we strengthen the radius bound
to $L(q)=O(\sqrt q)\,R$ for integers
$1\leq q\leq\lfloor n/2\rfloor+1$.
The proof compares the $q$th nearest site distance with the
distance at a smaller neighbor rank. At sites with sufficiently
many neighbors within a small fraction of the larger radius,
we obtain a stronger lower bound on $R$ using Lipschitz
functions. The remaining radii are bounded recursively at
the smaller rank. With the stronger radius bound, choosing
$q$ of order $1/t$ in the same recurrence gives
$H(t)=O(\sqrt t)\,R$, with a separate estimate when $q>n/2$.
We then pass from independent sampling to uniform subsets
of fixed size and use the symmetry between a subset and its
complement. We show that there is an absolute constant $C$
such that, for every $1\leq k\leq n$,
\[
    g_k\leq
    \frac{C}{n}\sqrt{\frac{n-k}{k}}\,\E[\OPT_\nu].
\]
This extends the bound of Gupta et al.~\cite[Lemma~4.5]{GGPW}
for tree metrics to arbitrary metric spaces, up to a constant
factor.

\paragraph{Roadmap.}
Section~\ref{sec:fair-bias} proves Theorem~\ref{thm:canonical} by
expressing the expected cost of Fair-Bias as a sampling integral,
grouping nearby servers, and deriving a recurrence for the sampling
cost. Section~\ref{sec:unknown-arrivals} proves Theorem~\ref{thm:unknown}
using Harada's reduction and projection to the nearest server.
Appendix~\ref{sec:sharp} gives a stronger bound on the expected cost
of a single match and shows that the exponent $1/2$ in the bound for
uniform subsets is optimal.
\section{Constant competitiveness of Fair-Bias}\label{sec:fair-bias}

We prove Theorem~\ref{thm:canonical} in this section. All requests
are drawn independently from the server distribution $\nu$, and
$\ALG$ denotes the total cost of Fair-Bias.

\subsection{The cost of a random subset}\label{sec:sampling}

Assume for now that $n\geq 2$ and that the servers occupy distinct
locations, which we call \emph{sites}. We identify each site with its
label, write $d(i,j)=d(s_i,s_j)$, and use $\delta_i$ for $\delta_{s_i}$. For every nonempty
$F\subseteq[n]$, define
\[
    \nu_F=\frac{1}{|F|}\sum_{i\in F}\delta_i,
    \qquad v_i=\delta_i-\nu\quad(i\in[n]).
\]
Since $\nu=n^{-1}\sum_{i=1}^n\delta_i$, each $v_i$ has total mass zero
and $\sum_{i=1}^n v_i=0$.

For a signed measure $z$ of total mass zero on the sites, let
$\tn{z}$ be the minimum cost of transporting its positive part to its
negative part. This is a norm on the space of signed measures of total
mass zero, since the sites are distinct. Homogeneity and symmetry
follow by scaling and reversing transport plans, respectively. The triangle inequality follows
by combining transport plans and shortcutting intermediate moves.
In particular, the norm is convex. For probability measures
$\lambda$ and $\eta$ on the sites, their transport cost is
$W_1(\lambda,\eta)=\tn{\lambda-\eta}$. For a real-valued function $f$
on the sites, write $\nu(f)=n^{-1}\sum_{i=1}^n f(i)$. Let $z(i)$
denote the mass of $z$ at site $i$, and define the pairing
$\langle z,f\rangle=\sum_{i=1}^n z(i)f(i)$. If $f$ is $1$-Lipschitz,
then moving one unit of mass from $i$ to $j$ changes this pairing by
at most $d(i,j)$. Thus
\[
    |\langle z,f\rangle|\leq\tn{z}.
\]

\paragraph{The cost of a single match.}
Let $F\subseteq[n]$ be the nonempty set of free server labels.
Before the next request arrives, Fair-Bias computes an optimal
transport plan $\rho_F$ from $\nu$ to $\nu_F$. Let $\rho_F(i,j)$ be
the mass transported from $i\in[n]$ to $j\in F$. We use the indexing of
Harada~\cite[Section~3.1]{Harada}; the transpose of $\rho_F$ solves
the fractional matching LP of Gupta et al.~\cite[Algorithm~1]{GGPW}.
Each row has mass $1/n$ and each column has mass $1/|F|$.
On a request at site $i$, Fair-Bias chooses server $j\in F$ with
probability $n\rho_F(i,j)$. Since the request has
law $\nu$, each free server is chosen with probability $1/|F|$, and
the expected cost of the match, given $F$, is $W_1(\nu_F,\nu)$.
For each $1\leq k\leq n$, let $F_k$ be the set of remaining server
labels when $k$ servers are free. By induction, $F_k$ is a uniformly
random $k$-subset of $[n]$~\cite[Lemma~2.2]{GGPW}. Following
Harada~\cite[Section~2.2]{Harada}, let
\[
    g_k=\E[W_1(\nu_{F_k},\nu)]
\]
be the expected cost of this match. The expected total cost of
Fair-Bias is $\sum_{k=1}^n g_k$.

\paragraph{Independent sampling.}
Giving each label in $F_k$ one unit of mass scales the transport cost
by $k$. Since $\sum_{i\in F_k}v_i=k(\nu_{F_k}-\nu)$, define
\[
    h_k=\E\!\left[\tn{\sum_{i\in F_k}v_i}\right]=kg_k
    \qquad (1\leq k\leq n),
\]
and put $h_0=0$. Note that $h_n=0$ as well, since $\sum_i v_i=0$.
For $t\in[0,1]$, sample each label independently with
probability $t$. Let $J_i(t)$ be the indicator that label $i$ is sampled,
and let $K(t)=\sum_{i=1}^n J_i(t)$ be the sample size. Give each sampled
label one unit of mass and redistribute the mass so that each of the
$n$ labels receives $K(t)/n$ units. The expected minimum cost is
\[
    H(t)=\E\!\left[\tn{\sum_{i=1}^n J_i(t)v_i}\right].
\]
Given the sample size, the sampled labels form a uniform subset of
that size. Thus the conditional expected transport cost given $K(t)$
is $h_{K(t)}$. Since $K(t)$ has the binomial distribution
$\operatorname{Bin}(n,t)$, we have
\[
    H(t)=\sum_{k=0}^n\binom{n}{k}t^k(1-t)^{n-k}h_k.
\]
Since $h_0=0$, we can divide by $t>0$ and integrate term by term.
For each $1\leq k\leq n$,
\[
    \binom{n}{k}\int_0^1 t^{k-1}(1-t)^{n-k}\,dt=\frac{1}{k}.
\]
Using $h_k=kg_k$, we obtain the identity
\begin{equation}\label{eq:integral}
    \E[\ALG]=\sum_{k=1}^n g_k=\int_0^1 H(t)\,\frac{dt}{t}.
\end{equation}

\paragraph{Comparison with the offline optimum.}
Let $I_1,\ldots,I_n$ be independent uniform labels in $[n]$, and let
$\OPT_\nu$ be the minimum perfect matching cost for the corresponding
requests. Each request places one unit of mass at its site, and each
server requires one unit. The resulting signed measure is
$\sum_{j=1}^n v_{I_j}$. Common mass at a site can be matched at zero
cost, and integer supplies and demands admit an optimal integral
transport plan. Hence
\[
    \E[\OPT_\nu]=\E\!\left[\tn{\sum_{j=1}^n v_{I_j}}\right].
\]
Let $\varepsilon_1,\ldots,\varepsilon_n$ be independent random variables,
each uniform on $\{-1,1\}$, and define
$R=\E\!\left[\tn{\sum_{i=1}^n\varepsilon_i v_i}\right]$.
We will use the following bounds for every $i\in[n]$:
\begin{equation}\label{eq:bench}
    \tn{v_i}\leq R=2H(1/2)\leq 2\E[\OPT_\nu].
\end{equation}
The first inequality follows by conditioning on $\varepsilon_i$ and
applying Jensen's inequality. The equality follows from writing
$\varepsilon_i=2J_i(1/2)-1$ under a coupling and using $\sum_i v_i=0$.

For the last inequality, we compare sampling with and without
replacement~\cite[Lemma~4.3]{GGPW}. Fix $1\leq k\leq n$.
Extend the set of distinct labels among $I_1,\ldots,I_k$ to a
$k$-subset $U_k$ of $[n]$ by choosing the additional labels uniformly
from its complement. By symmetry under permutations of $[n]$, the
set $U_k$ is a uniform $k$-subset. Conditional on $U_k$, the
multiplicities of its labels are exchangeable by symmetry under
permutations within $U_k$, and they sum to $k$.
Each label in $U_k$ therefore occurs once in expectation, so
\[
    \E\!\left[\sum_{j=1}^k v_{I_j}\,\middle|\,U_k\right]
    =\sum_{i\in U_k}v_i.
\]
Since $U_k$ has the same distribution as $F_k$, Jensen's inequality
gives
\[
    h_k\leq\E\!\left[\tn{\sum_{j=1}^k v_{I_j}}\right]
    \leq\E[\OPT_\nu].
\]
The second inequality follows by adding the remaining independent
terms $v_{I_{k+1}},\ldots,v_{I_n}$, each of which has mean
$n^{-1}\sum_{i=1}^n v_i=0$, and applying Jensen's inequality again.
Averaging over the sample size at $t=1/2$ gives
$H(1/2)\leq\E[\OPT_\nu]$, completing the proof of~\eqref{eq:bench}.
The bound $g_k=h_k/k\leq\E[\OPT_\nu]/k$ alone gives only an
$O(\log n)$ competitive ratio. We instead use~\eqref{eq:integral}
and bound $H(t)$ as $t$ tends to zero.

\subsection{Grouping nearby servers}\label{sec:geometry}

We continue to assume that the server sites are distinct. For an integer
$1\leq q\leq n$ and a site $i\in[n]$, let $r_i(q)$ be the distance
from $i$ to its $q$th nearest site, counting $i$ itself. Equivalently,
$r_i(q)$ is the smallest radius for which the closed ball about $i$
contains at least $q$ sites. Write $L(q)=\sum_{i=1}^n r_i(q)$.
We first bound $L(q)$ in terms of $R$ and then use this bound to
control the cost of transporting mass within groups of nearby sites.

\begin{lemma}\label{lem:radii}
For every integer $2\leq q\leq \lfloor n/2\rfloor$,
\[
    L(q)\leq 8qR.
\]
\end{lemma}

\begin{proof}
Fix $q$ in the stated range and write $r_i=r_i(q)$ for each $i\in[n]$.
Every site within distance $r_i$ of $i$ lies within distance
$r_i+d(i,j)$ of $j$. Since there are at least $q$ such sites,
$r_j\leq r_i+d(i,j)$. By symmetry, $|r_i-r_j|\leq d(i,j)$.

Construct a graph on $[n]$ by joining distinct sites $i$ and $j$ if
$d(i,j)<(r_i+r_j)/4$. For every neighbor $j$ of $i$, the inequality
$r_j\leq r_i+d(i,j)$ gives $d(i,j)<2r_i/3$. The open ball of radius
$r_i$ about $i$ contains fewer than $q$ sites, including $i$ itself.
Thus the graph has maximum degree at most $q-2$ and can be greedily
colored with at most $q-1$ colors. Some color class $S$ satisfies
$\sum_{i\in S}r_i\geq L(q)/q$.

For each $i\in S$, define $f_i:[n]\to\mathbb R$ by
$f_i(x)=\max\{r_i/4-d(i,x),0\}$. Since $S$ is an independent set,
$d(i,j)\geq(r_i+r_j)/4$ for distinct $i,j\in S$. The positive supports
of the functions $f_i$ are therefore disjoint. Moreover, for any signs
$\sigma_i\in\{-1,1\}$, the function $\sum_{i\in S}\sigma_i f_i$ is
$1$-Lipschitz. Indeed, if $x$ and $y$ lie in the positive supports of
$f_i$ and $f_j$, respectively, with $i\neq j$, then
\[
    f_i(x)+f_j(y)
    \leq d(i,j)-d(i,x)-d(j,y)
    \leq d(x,y).
\]
If both points lie in the same support, or only one lies in a support,
the claim follows from the Lipschitz property of a single $f_i$.
If neither point lies in a support, the signed sum vanishes at both.

Recall that $R=\E\Bigl[\tn{\sum_{j=1}^n\varepsilon_jv_j}\Bigr]$, where the
$\varepsilon_j$ are independent uniform signs. For each realization
of these signs, take $f=\sum_{i\in S}\varepsilon_i f_i$. Since $f$ is
$1$-Lipschitz and the products of distinct signs have mean zero,
\[
    R\geq
    \E\!\left[\left\langle\sum_{j=1}^n\varepsilon_jv_j,f
    \right\rangle\right]
    =\sum_{i\in S}\bigl(f_i(i)-\nu(f_i)\bigr).
\]
Each $f_i$ is positive on fewer than $q\leq n/2$ sites and is bounded
above by $f_i(i)=r_i/4$. Hence $\nu(f_i)\leq r_i/8$, and
\[
    R\geq\frac18\sum_{i\in S}r_i\geq\frac{L(q)}{8q}.
\]
This proves the lemma.
\end{proof}

\begin{lemma}\label{lem:groups}
For every integer $2\leq q\leq\lfloor n/2\rfloor$, there is a
partition $\mathcal G$ of $[n]$ such that $q\leq |G|\leq 2q$ for every
$G\in\mathcal G$ and
\[
    \sum_{G\in\mathcal G}\sum_{i\in G}\tn{\delta_i-\nu_G}
    \leq 4L(q).
\]
\end{lemma}

\begin{proof}
Write $r_i=r_i(q)$. Process the sites in nondecreasing order of $r_i$,
breaking ties arbitrarily. Select $i$ as a center if no previously
selected center is within distance $2r_i$ of $i$. Let $\mathcal C$ be
the resulting set of centers. The closed balls of radii $r_c$ about
$c\in\mathcal C$ are pairwise disjoint. Indeed, if $c'$ was selected
before $c$, then $r_{c'}\leq r_c$ and
$d(c,c')>2r_c\geq r_c+r_{c'}$. Each ball contains at least $q$ sites.

Assign every site in one of these balls to its center. Assign every
remaining site $i$ to a center within distance $2r_i$, which exists
by the selection rule. We claim that every site $i$ is within distance
$2r_i$ of its assigned center. It suffices to consider a site $i$ in
the ball about a center $c$ with $r_i<r_c$. When $i$ was processed,
either $i$ itself was selected as a center or an earlier center $c'$
was within distance $2r_i$ of $i$. In either case, there is a center
$c'$ selected before $c$ with $d(i,c')\leq 2r_i$. Consequently,
\[
    2r_c<d(c,c')\leq d(c,i)+d(i,c')\leq r_c+2r_i.
\]
Thus $r_c<2r_i$ and $d(i,c)\leq r_c<2r_i$.

The sites assigned to each center form a group of at least $q$ sites.
Split each group of more than $2q$ sites into parts of sizes between
$q$ and $2q$. Such a split is always possible by taking parts of size
$q$ and adding any remaining sites to the last part. Let $\mathcal G$
be the resulting partition. Each part $G\in\mathcal G$ retains the
center $c$ of its original group as a reference point, even if $c$ is
not in $G$. Recall that $\nu_G$ is the uniform distribution on $G$.
For each $i\in G$, we have
\[
    \tn{\delta_i-\nu_G}
    =\frac{1}{|G|}\sum_{j\in G}d(i,j)
    \leq d(i,c)+\frac{1}{|G|}\sum_{j\in G}d(c,j).
\]
Summing over $i\in G$ and using $d(i,c)\leq 2r_i$ gives
\[
    \sum_{i\in G}\tn{\delta_i-\nu_G}
    \leq 2\sum_{i\in G}d(i,c)
    \leq 4\sum_{i\in G}r_i.
\]
Summing over all $G\in\mathcal G$ proves the lemma.
\end{proof}

\subsection{A recurrence for the sampling cost}\label{sec:counts}

We continue to assume that the server sites are distinct. Fix a sampling
probability $0<t\leq 1$ and an integer $2\leq q\leq\lfloor n/2\rfloor$
with $tq\leq 2$. Let $\mathcal G$ be a partition as in
Lemma~\ref{lem:groups}, chosen before sampling and kept fixed throughout
the proof. Each group contains between $q$ and $2q$ sites,
so its expected number of sampled sites is at most four. After spreading
the mass at each sampled site uniformly over its group, the remaining
transport cost depends only on the number of sampled sites in each group.

\begin{proposition}\label{prop:recurrence}
For every $0<t\leq 1$ and integer $2\leq q\leq\lfloor n/2\rfloor$
with $tq\leq 2$, we have
\begin{equation}\label{eq:recurrence}
    H(t)\leq 4tL(q)+24\sqrt{3/q}\,R
        +2\sum_{\ell=25}^{\infty} H(c_\ell t),
    \qquad c_\ell=(4e/\ell)^\ell.
\end{equation}
\end{proposition}

\begin{proof}
For each group $G\in\mathcal G$, let $w_G=\nu_G-\nu$ and let
$K_G=\sum_{i\in G}J_i(t)$ be its sample count. The counts are independent,
with $K_G\sim\operatorname{Bin}(|G|,t)$. For each $i\in G$, write
$v_i=(\delta_i-\nu_G)+w_G$. By the triangle inequality and
Lemma~\ref{lem:groups}, the expected cost of the first part is at most
\[
    t\sum_{G\in\mathcal G}\sum_{i\in G}\tn{\delta_i-\nu_G}
    \leq 4tL(q).
\]
Since $\sum_{G\in\mathcal G}|G|w_G=0$, the expected cost of the
remaining signed measure is
\begin{equation}\label{eq:countcost}
    \E\!\left[\tn{\sum_{G\in\mathcal G}(K_G-t|G|)w_G}\right].
\end{equation}
We write $\mathbf{1}_{\mathcal E}$ for the indicator of an event $\mathcal E$. Let
$\bar K_G=\min\{K_G,24\}$. Since $\E[K_G]=t|G|$, we have
\[
    K_G-t|G|=(\bar K_G-\E[\bar K_G])
        +\sum_{\ell\geq 25}
        \bigl(\mathbf{1}_{\{K_G\geq\ell\}}-\Pr(K_G\geq\ell)\bigr).
\]
The sum has only finitely many nonzero terms, since $K_G\leq |G|$.
The cutoff at $24$ ensures that $4e/\ell<1/2$ for every $\ell\geq 25$.
We bound the two parts separately using the triangle inequality.
Throughout the proof, sums indexed by $G$ run over $\mathcal G$.

\paragraph{Counts up to 24.}
Let $(\bar K'_G)_G$ be an independent copy of $(\bar K_G)_G$.
Conditional on $(\bar K_G)_G$, the difference $\bar K_G-\bar K'_G$
has mean $\bar K_G-\E[\bar K_G]$. By Jensen's inequality,
\[
    \E\!\left[\tn{\sum_G(\bar K_G-\E[\bar K_G])w_G}\right]
    \leq \E\!\left[\tn{\sum_G(\bar K_G-\bar K'_G)w_G}\right].
\]
The differences are independent, symmetric about zero, and bounded in
absolute value by $24$. Let $(\xi_G)_G$ be independent uniform signs,
independent of the differences. Conditional on their absolute values,
the differences have the joint distribution of
$(\xi_G|\bar K_G-\bar K'_G|)_G$. For any fixed signed measure $z$ of
mass zero, the function
\[
    a\longmapsto\frac{\tn{z+aw_G}+\tn{z-aw_G}}{2}
\]
is even and convex, and hence nondecreasing for $a\geq 0$. Averaging
over one sign at a time therefore gives
\begin{equation}\label{eq:truncatedcounts}
    \E\!\left[\tn{\sum_G(\bar K_G-\E[\bar K_G])w_G}\right]
    \leq 24\,\E\!\left[\tn{\sum_G\xi_G w_G}\right].
\end{equation}

To relate the group signs to the site signs in $R$, let
$T_G=\sum_{i\in G}\varepsilon_i$. By symmetry within each group and
independence across groups, the conditional mean of $\varepsilon_i$
given all the group totals is $T_G/|G|$ for $i\in G$. Since
$\sum_{i\in G}v_i=|G|w_G$, Jensen's inequality gives
\[
    R\geq\E\!\left[\tn{\sum_G T_Gw_G}\right].
\]
The second and fourth moments satisfy $\E[T_G^2]=|G|$ and
$\E[T_G^4]=3|G|^2-2|G|\leq 3|G|^2$. By H\"older's inequality,
\[
    \E[|T_G|]\geq
    \frac{(\E[T_G^2])^{3/2}}{(\E[T_G^4])^{1/2}}
    \geq\sqrt{|G|/3}\geq\sqrt{q/3}.
\]
The totals are independent and symmetric about zero, so $(T_G)_G$
has the joint distribution of $(\xi_G|T_G|)_G$, with the signs chosen
independently of the absolute values. Applying Jensen's inequality
over the absolute values, followed by the same monotonicity, gives
\[
    \begin{aligned}
    R&\geq\E\!\left[\tn{\sum_G T_Gw_G}\right]\\
     &\geq\E\!\left[\tn{\sum_G\xi_G\,\E[|T_G|]\,w_G}\right]
      \geq\sqrt{q/3}\,\E\!\left[\tn{\sum_G\xi_Gw_G}\right].
    \end{aligned}
\]
Together with~\eqref{eq:truncatedcounts}, this bounds the contribution
of the truncated counts by $24\sqrt{3/q}\,R$.

\paragraph{Counts above 24.}
Fix an integer $\ell\geq 25$ and set $u=c_\ell t$. Take a fresh sample
by selecting each site independently with probability $u$, independently
of the original sample. Let $\widetilde J_i(u)$ indicate whether site
$i$ is selected in the fresh sample, and let
$Y_G=\sum_{i\in G}\widetilde J_i(u)$. Thus the $Y_G$ are independent,
with $Y_G\sim\operatorname{Bin}(|G|,u)$. We show that the contribution
of level $\ell$ to~\eqref{eq:countcost} is at most $2H(u)$.

Fix a group $G$ and write $p=\Pr(K_G\geq\ell)$ and
$p'=\Pr(Y_G\geq 1)$. If $\ell>|G|$, then $p=0$. Otherwise, a union
bound over all choices of $\ell$ sites gives
$p\leq\binom{|G|}{\ell}t^\ell\leq(e|G|t/\ell)^\ell$. Since
$|G|\leq 2q$ and $qt\leq 2$, in either case we have
\[
    p\leq (2eqt/\ell)^\ell
      =c_\ell(qt/2)^\ell\leq c_\ell qt/2.
\]
Also, $u|G|\leq 4c_\ell<1/2$ since $c_\ell<2^{-\ell}$. Using
$|G|\geq q$, the identity $p'=1-(1-u)^{|G|}$, and the inequality
$1-e^{-x}\geq x/2$ for $0\leq x\leq 1$, we obtain
\[
    p\leq u|G|/2\leq 1-e^{-u|G|}\leq p'\leq u|G|<1/2.
\]
The bound $p'\leq u|G|$ follows from a union bound.

We claim that for every convex function $\varphi:\mathbb R\to\mathbb R$,
\begin{equation}\label{eq:onegroup}
    \E\!\left[\varphi(\mathbf{1}_{\{K_G\geq\ell\}}-p)\right]
    \leq \E\!\left[\varphi\bigl(2(Y_G-u|G|)\bigr)\right].
\end{equation}
To prove this, condition $2(Y_G-u|G|)$ on whether $Y_G\geq 1$.
Its conditional mean takes two values:
\[
    a=-2u|G|\quad\text{if }Y_G=0,\qquad
    b=\frac{2u|G|}{p'}(1-p')\quad\text{if }Y_G\geq 1.
\]
Here $p'>0$ and $\E[Y_G\mid Y_G\geq 1]=u|G|/p'$.
The bounds above imply $a\leq-p$ and $b\geq 1\geq 1-p$.
Thus the centered indicator takes values in $[a,b]$, and both it and
the conditional mean have mean zero. The chord of $\varphi$ on $[a,b]$
lies above $\varphi$ and agrees with it at the endpoints. Since the
chord is affine, its expectation is the same for both random variables.
The expected value of $\varphi$ at the centered indicator is therefore
at most its expected value at the conditional mean. Applying Jensen's
inequality to the conditional mean proves~\eqref{eq:onegroup}.

The transport norm is convex, so for any fixed signed measure $z$ of
total mass zero, the function $x\mapsto\tn{z+xw_G}$ is convex.
We can thus apply~\eqref{eq:onegroup} to the contribution of group $G$,
conditioning on the sum of the contributions of all other groups.
At each replacement the other groups are independent of both coefficients
being compared. Replacing the coefficients one group at a time and
using homogeneity of the norm gives
\[
    \E\!\left[\tn{\sum_G
        (\mathbf{1}_{\{K_G\geq\ell\}}-\Pr(K_G\geq\ell))w_G}\right]
    \leq 2\,\E\!\left[\tn{\sum_G(Y_G-u|G|)w_G}\right].
\]
Given its group counts, the fresh sample is uniform within each group.
Using $\sum_{i\in G}v_i=|G|w_G$ and $\sum_G|G|w_G=0$, we have
\[
    \E\!\left[\sum_{i=1}^n\widetilde J_i(u)v_i\,\middle|\,(Y_G)_G\right]
    =\sum_G Y_Gw_G=\sum_G(Y_G-u|G|)w_G.
\]
By Jensen's inequality, the expected norm on the right is at most
$H(u)$. Hence level $\ell$ contributes at most $2H(c_\ell t)$.
The triangle inequality allows us to sum these bounds over all levels,
without requiring independence between levels. Combining them with
the bounds for the truncated counts and the transport within groups
proves~\eqref{eq:recurrence}.
\end{proof}

\subsection{The competitive ratio}\label{sec:competitive}

We first bound $H(t)$ for distinct server sites. By Lemma~\ref{lem:radii},
the first two terms in~\eqref{eq:recurrence} are at most $32tqR$ and
$24\sqrt{3/q}\,R$, respectively. Balancing $tq$ and $q^{-1/2}$ suggests
choosing $q$ of order $t^{-2/3}$.

\begin{lemma}\label{lem:density}
For every $0<t\leq 1$,
\[
    H(t)\leq 114\,t^{1/3}R.
\]
\end{lemma}

\begin{proof}
Fix $0<t\leq 1$ and choose $q=\max\{2,\lceil t^{-2/3}\rceil\}$. Then
$t^{-2/3}\leq q\leq 2t^{-2/3}$ and $tq\leq 2$. By the triangle inequality
and~\eqref{eq:bench},
\[
    H(t)\leq t\sum_{i=1}^n\tn{v_i}\leq ntR.
\]
If $q>n/2$, then $n<2q\leq 4t^{-2/3}$, so $H(t)\leq 4t^{1/3}R$.
Otherwise, Proposition~\ref{prop:recurrence} applies, and the first two
terms in~\eqref{eq:recurrence} sum to at most
$(64+24\sqrt{3})t^{1/3}R$.

Fix the server placement and define
\[
    M_{1/3}=\sup_{0<u\leq 1}\frac{H(u)}{u^{1/3}R}.
\]
Since $n\geq 2$ and the sites are distinct, $R\geq\tn{v_i}>0$ by
\eqref{eq:bench}. The bound $H(u)\leq nuR$ for every $0<u\leq 1$
therefore implies $M_{1/3}\leq n$. Recall that
$c_\ell=(4e/\ell)^\ell$ for integers $\ell\geq 25$. Since
$4e/\ell<1/2$, we have $0<c_\ell<1$ and
$H(c_\ell t)\leq M_{1/3}c_\ell^{1/3}t^{1/3}R$. Moreover,
\begin{equation}\label{eq:tailseries}
    2\sum_{\ell\geq 25}c_\ell^{1/3}
    \leq 2\sum_{\ell\geq 25}2^{-\ell/3}
    \leq 6\sum_{j\geq 8}2^{-j}
    =\frac{3}{64}<\frac{1}{16},
\end{equation}
where the second inequality groups terms according to
$j=\lfloor\ell/3\rfloor$. Combining the two cases for $q$, dividing by
$t^{1/3}R$, and taking the supremum over $t$ gives
\[
    M_{1/3}\leq\max\left\{4,64+24\sqrt{3}+\frac{M_{1/3}}{16}\right\}.
\]
Hence $M_{1/3}\leq\frac{16}{15}(64+24\sqrt{3})<114$.
\end{proof}

\begin{proof}[Proof of Theorem~\ref{thm:canonical}]
First suppose that $n\geq 2$ and the server sites are distinct.
By~\eqref{eq:integral}, Lemma~\ref{lem:density}, and~\eqref{eq:bench},
\[
    \E[\ALG]=\sum_{k=1}^n g_k
    \leq 114R\int_0^1 t^{-2/3}\,dt
    =342R\leq 684\,\E[\OPT_\nu]\leq 700\,\E[\OPT_\nu].
\]
Rounding up gives the $700$-competitive bound for distinct sites.

Now allow several server labels to share a location. For each request,
choose a label uniformly from all original labels at its location,
independently of the other choices. Since requests are drawn independently
from $\nu$, the resulting labels are independent and uniform on $[n]$.
Fair-Bias still uses the plan $\rho_F$ indexed by server labels, with
cost $d(s_i,s_j)$ between labels $i$ and $j$. Each row has mass $1/n$ and
each free label has column mass $1/|F|$. Thus the free set is uniform at
each size, and the expected total cost is still $\sum_{k=1}^n g_k$,
where $g_k$ is defined using these transport costs.

For $\epsilon>0$, define a metric on $[n]$ by
\[
    d_\epsilon(i,j)=
    \begin{cases}
        0, & i=j,\\
        d(s_i,s_j)+\epsilon, & i\neq j.
    \end{cases}
\]
The triangle inequality follows from that of $d$, and distinct labels
now have positive distance. Let $g_k^\epsilon$ and $\OPT_\nu^\epsilon$
be the corresponding expected cost of a single match and the offline
optimum for the same request labels under $d_\epsilon$, respectively.
For each fixed free set, the feasible transport plans are unchanged and
have total mass one. Their costs increase by at most $\epsilon$ and
cannot decrease. Averaging over the uniform free set gives
$0\leq g_k^\epsilon-g_k\leq\epsilon$. Likewise, each perfect matching
uses $n$ edges, so $0\leq\OPT_\nu^\epsilon-\OPT_\nu\leq n\epsilon$
for every realization of the requests. Applying the distinct-site bound
to $d_\epsilon$ gives
\[
    \sum_{k=1}^n g_k
    \leq\sum_{k=1}^n g_k^\epsilon
    \leq 700\,\E[\OPT_\nu^\epsilon]
    \leq 700\,\E[\OPT_\nu]+700n\epsilon.
\]
Letting $\epsilon$ tend to zero proves the result. If $n=1$, every
request drawn from $\nu$ is at the unique server, so both costs are zero.
\end{proof}

\section{Unknown i.i.d.\ arrivals}\label{sec:unknown-arrivals}

\begin{proof}[Proof of Theorem~\ref{thm:unknown}]
On a finite metric space, Harada's algorithm~\cite[Algorithm~1]{Harada}
computes an optimal transport plan from the empirical distribution of
all requests seen so far, including the current request, to the server
distribution $\nu$. The plan is chosen as a function of the request
multiset, with one column per server label. Equal request locations
can be represented by a single row, or by identical rows of equal mass.
The row corresponding to the current request is normalized to a
probability distribution, from which a virtual server label is sampled. The algorithm gives this virtual
request to Fair-Bias and matches the actual request to the server chosen
by Fair-Bias. The randomness used to sample the virtual labels is fresh
at each step and independent of the request sequence and of Fair-Bias's
randomness. Fair-Bias uses only the virtual labels, its free set, and its
own random choices. Harada~\cite[Theorem~6.2]{Harada} showed that if Fair-Bias
is $\alpha$-competitive for independent requests from $\nu$, then his
algorithm is $(1+6\alpha)$-competitive for unknown i.i.d.\ arrivals on
finite metric spaces, including requests at locations without servers.
Theorem~\ref{thm:canonical} gives a competitive ratio of
$1+6\cdot 700=4201$.

For an arbitrary metric space, we use the projection described
in~\cite{Harada,MNP06}. Let $\pi(x)$ be a nearest server location to a
request at $x\in X$, chosen from all original servers, with ties broken
by a fixed order of the labels. Run the finite metric space algorithm
on the projected requests and match each actual request to the server
it selects. The preimage of each server location under $\pi$ is Borel,
since the choice can be described by finitely many comparisons of
continuous distance functions. The projected requests are therefore
again i.i.d.

Let $x_1,\ldots,x_n$ be the realized requests and let
$P=\sum_{j=1}^n d(x_j,\pi(x_j))$ be the total projection cost.
Write $\widehat{\OPT}$ and $\widehat{\ALG}$ for the offline optimum and
the online cost on the projected requests, respectively. Each request
is at least as close to its nearest server as to its server in the
original offline matching, so $P\leq\OPT$. By the triangle inequality,
using the original offline matching for the projected requests gives
$\widehat{\OPT}\leq\OPT+P\leq 2\OPT$. The same inequality gives
$\ALG\leq\widehat{\ALG}+P$. Taking expectations yields
\[
    \E[\ALG]\leq 4201\,\E[\widehat{\OPT}]+\E[P]
    \leq (2\cdot 4201+1)\E[\OPT]=8403\,\E[\OPT].
\]
This completes the proof.
\end{proof}

\section*{AI methodologies}
\phantomsection\label{ai-usage}
\addcontentsline{toc}{section}{AI methodologies}
We used ChatGPT (GPT-6.2) to explore some of the ideas
developed in this paper. We also used it to develop,
check, and refine portions of the proofs, search the
literature, and edit the manuscript and its \LaTeX{} source.

\bibliographystyle{abbrv}
\bibliography{references}

\begin{thebibliography}{10}

\bibitem{BFP23}
E.~Balkanski, Y.~Faenza, and N.~P{\'e}rivier.
\newblock The power of greedy for online minimum cost matching on the line.
\newblock In K.~Leyton-Brown, J.~D. Hartline, and L.~Samuelson, editors, {\em Proceedings of the 24th {ACM} Conference on Economics and Computation ({EC})}, pages 185--205. {ACM}, 2023.

\bibitem{BBGN14}
N.~Bansal, N.~Buchbinder, A.~Gupta, and J.~Naor.
\newblock A randomized {$O(\log^2 k)$}-competitive algorithm for metric bipartite matching.
\newblock {\em Algorithmica}, 68(2):390--403, 2014.

\bibitem{FRT04}
J.~Fakcharoenphol, S.~Rao, and K.~Talwar.
\newblock A tight bound on approximating arbitrary metrics by tree metrics.
\newblock {\em Journal of Computer and System Sciences}, 69(3):485--497, 2004.

\bibitem{GGPW}
A.~Gupta, G.~Guruganesh, B.~Peng, and D.~Wajc.
\newblock Stochastic online metric matching.
\newblock In {\em 46th International Colloquium on Automata, Languages, and Programming ({ICALP})}, volume 132 of {\em LIPIcs}, pages 67:1--67:14. Schloss Dagstuhl - Leibniz-Zentrum f{\"u}r Informatik, 2019.
\newblock Full version: arXiv:1904.09284v1.

\bibitem{GL12}
A.~Gupta and K.~Lewi.
\newblock The online metric matching problem for doubling metrics.
\newblock In A.~Czumaj, K.~Mehlhorn, A.~M. Pitts, and R.~Wattenhofer, editors, {\em Automata, Languages, and Programming ({ICALP}), Part I}, volume 7391 of {\em Lecture Notes in Computer Science}, pages 424--435. Springer, 2012.

\bibitem{Harada}
T.~Harada.
\newblock Online metric matching with unknown {i.i.d.} arrivals.
\newblock {\em arXiv preprint arXiv:2611.04275}, 2026.

\bibitem{KP93}
B.~Kalyanasundaram and K.~Pruhs.
\newblock Online weighted matching.
\newblock {\em Journal of Algorithms}, 14(3):478--488, 1993.

\bibitem{KMV94}
S.~Khuller, S.~G. Mitchell, and V.~V. Vazirani.
\newblock On-line algorithms for weighted bipartite matching and stable marriages.
\newblock {\em Theoretical Computer Science}, 127(2):255--267, 1994.

\bibitem{LVY26}
Y.~Li, E.~Vitercik, and M.~Yang.
\newblock Smoothed analysis of online metric matching with a single sample: Beyond metric distortion.
\newblock In {\em 17th Innovations in Theoretical Computer Science Conference, {ITCS} 2026, Bocconi University, Milan, Italy, January 27-30, 2026}, volume 362 of {\em LIPIcs}, pages 94:1--94:23. Schloss Dagstuhl - Leibniz-Zentrum f{\"{u}}r Informatik, 2026.

\bibitem{MNP06}
A.~Meyerson, A.~Nanavati, and L.~J. Poplawski.
\newblock Randomized online algorithms for minimum metric bipartite matching.
\newblock In {\em Proceedings of the Seventeenth Annual {ACM-SIAM} Symposium on Discrete Algorithms ({SODA})}, pages 954--959. {ACM} Press, 2006.

\bibitem{Rag16}
S.~Raghvendra.
\newblock A robust and optimal online algorithm for minimum metric bipartite matching.
\newblock In {\em Approximation, Randomization, and Combinatorial Optimization. Algorithms and Techniques ({APPROX/RANDOM})}, volume~60 of {\em LIPIcs}, pages 18:1--18:16. Schloss Dagstuhl - Leibniz-Zentrum f{\"u}r Informatik, 2016.

\bibitem{SYY24}
A.~Saberi, M.~Yang, and S.~H. Yu.
\newblock Stochastic online metric matching: Adversarial is no harder than stochastic.
\newblock {\em CoRR}, abs/2407.14785v1, 2024.
\newblock Also in {WINE} 2024.

\bibitem{YY26}
M.~Yang and S.~H. Yu.
\newblock Online metric matching: Beyond the worst case.
\newblock {\em Oper. Res.}, 74(1):130--140, 2026.

\bibitem{YYGreedy26}
M.~Yang and S.~H. Yu.
\newblock Optimal analysis of {Greedy} for stochastic online {Euclidean} matching: Flat torus to the rescue.
\newblock {\em arXiv preprint arXiv:2609.23947}, 2026.

\end{thebibliography}

\appendix
\numberwithin{equation}{section}
\section{The cost of a single match}\label{sec:sharp}

Gupta et al.~\cite[Lemma~4.5]{GGPW} bounded the expected cost of a
single Fair-Bias match on tree metrics by summing the imbalances across
the tree edges. In our notation, their bound is
$g_k\leq 2\E[\OPT_\nu]/\sqrt{nk}$. We extend this bound to arbitrary
metrics, up to an absolute constant. Using the symmetry between a free
set and its complement also gives a factor of $\sqrt{(n-k)/n}$.

\begin{theorem}\label{thm:sharp-stage}
There is an absolute constant $C$ such that, for every $1\leq k\leq n$,
the expected cost of the next Fair-Bias match when $k$ servers are free
satisfies
\[
    g_k\leq\frac{C}{n}\sqrt{\frac{n-k}{k}}\,\E[\OPT_\nu],
\]
provided that the requests are drawn independently from the server
distribution $\nu$. The expectation in $g_k$ includes the randomness of
the free set $F_k$.
\end{theorem}

We first assume that $n\geq 2$ and the server sites are distinct.
Recall from Subsection~\ref{sec:sampling} that $v_i=\delta_i-\nu$,
$R=\E[\tn{\sum_i\varepsilon_i v_i}]$ for independent uniform signs
$\varepsilon_i$, and $H(t)=\E[\tn{\sum_i J_i(t)v_i}]$ for independent
Bernoulli sampling with probability $t$. We first show that there is an
absolute constant $C_1$ such that
\begin{equation}\label{eq:sharp-bernoulli}
    H(t)\leq C_1\sqrt t\,R\qquad (0<t\leq 1).
\end{equation}
Integrating this bound in~\eqref{eq:integral} also gives a constant
competitive ratio. The simpler radius bound of Lemma~\ref{lem:radii}
suffices for the explicit ratio of $700$. Here we strengthen that bound
and then use~\eqref{eq:sharp-bernoulli} to analyze subsets of fixed size.
Recall that $r_i(q)$ is the distance from site $i$ to its $q$th nearest
site, counting itself, and $L(q)=\sum_i r_i(q)$. In particular, $L(1)=0$
and $r_i(q)>0$ for $2\leq q\leq n$.

\begin{lemma}\label{lem:sharp-radii}
There is an absolute constant $C_0$ such that, for every integer
$1\leq q\leq\lfloor n/2\rfloor+1$,
\[
    L(q)\leq C_0\sqrt q\,R.
\]
\end{lemma}

\begin{proof}
The case $q=1$ follows from $L(1)=0$. Fix an integer
$2\leq q\leq\lfloor n/2\rfloor+1$, put
$q'=\max\{1,\lfloor q/256\rfloor\}$, and write $r_i=r_i(q)$.
Call a site $i$ good if $r_i(q')\leq r_i/8$. Process the good sites in
nonincreasing order of $r_i$, breaking ties by label. Retain $i$ as a
center if $d(i,c)>(r_i+r_c)/4$ for every previously retained center $c$.
Otherwise, assign $i$ to a center for which this inequality fails.
Let $\mathcal C$ be the set of retained centers, and assign each center
to itself.

\emph{Packing and charging.}
Every site $i$ assigned to a center $c$ satisfies $r_i\leq r_c$ and
$d(i,c)\leq (r_i+r_c)/4\leq r_c/2$. The closed ball of radius $r_c/2$
about $c$ contains fewer than $q$ sites, since $r_c>0$ is the distance
to the $q$th nearest site. The sites assigned to $c$ therefore contribute
at most $qr_c$ to the sum of radii. Summing over the centers gives
\begin{equation}\label{eq:sharp-packing}
    \sum_{i:\,i\ \mathrm{good}}r_i\leq q\sum_{c\in\mathcal C}r_c.
\end{equation}

\emph{Lipschitz witnesses.}
For each $c\in\mathcal C$, define $f_c:[n]\to\mathbb R$ by
$f_c(x)=\max\{r_c/4-d(c,x),0\}$. Distinct centers $c,c'$ satisfy
$d(c,c')>(r_c+r_{c'})/4$, so the positive supports of the functions are
disjoint. If $f_c(x)>0$ and $f_{c'}(y)>0$ for distinct centers $c,c'$, then
\[
    d(x,y)\geq d(c,c')-d(c,x)-d(c',y)>f_c(x)+f_{c'}(y).
\]
Thus $f=\sum_{c\in\mathcal C}\sigma_c f_c$ is $1$-Lipschitz for any
choice of signs $\sigma_c\in\{-1,1\}$. The inequality above handles
points in different positive supports. If both points belong to the
same support, or one point lies outside all supports, the claim follows
from the Lipschitz property of the corresponding $f_c$. If both points
lie outside all supports, $f$ vanishes at both.

\emph{A lower bound on $R$.}
Let $Z=\sum_i\varepsilon_i v_i$, so $R=\E[\tn{Z}]$. For a fixed center
$c\in\mathcal C$, define $a_{c,i}=f_c(i)-\nu(f_c)$ for $i\in[n]$ and
write $S_c=\langle Z,f_c\rangle=\sum_i a_{c,i}\varepsilon_i$. The positive
support of $f_c$ contains at most $q-1\leq\lfloor n/2\rfloor$ sites.
By Cauchy--Schwarz on this support,
\begin{equation}\label{eq:sharp-centering}
    \sum_i a_{c,i}^2
    =\sum_i f_c(i)^2-\frac1n\Bigl(\sum_i f_c(i)\Bigr)^2
    \geq\frac12\sum_i f_c(i)^2.
\end{equation}
Since $c$ is good, at least $q'$ sites lie within distance $r_c/8$ of $c$,
and $f_c(i)\geq r_c/8$ at each of them. Hence
\begin{equation}\label{eq:sharp-variance}
    \sum_i a_{c,i}^2\geq\frac{q'r_c^2}{128}.
\end{equation}
By independence of the signs,
\[
    \E[S_c^2]=\sum_i a_{c,i}^2,\qquad
    \E[S_c^4]=3\Bigl(\sum_i a_{c,i}^2\Bigr)^2-2\sum_i a_{c,i}^4
    \leq 3(\E[S_c^2])^2.
\]
H\"older's inequality gives
$\E[S_c^2]\leq (\E[|S_c|])^{2/3}(\E[S_c^4])^{1/3}$.
Since $\E[S_c^2]>0$ by~\eqref{eq:sharp-variance}, we obtain
\[
    \E[|S_c|]\geq\frac{(\E[S_c^2])^{3/2}}{(\E[S_c^4])^{1/2}}
    \geq\frac1{\sqrt3}\Bigl(\sum_i a_{c,i}^2\Bigr)^{1/2}.
\]
Combining this with~\eqref{eq:sharp-variance} yields
\begin{equation}\label{eq:sharp-pairing}
    \E[|\langle Z,f_c\rangle|]\geq\frac{\sqrt{q'}\,r_c}{8\sqrt6}.
\end{equation}
For each realization of $Z$, choose $\sigma_c$ to be the sign of
$\langle Z,f_c\rangle$, with either choice when the pairing is zero.
The resulting $f=\sum_{c\in\mathcal C}\sigma_c f_c$ is $1$-Lipschitz, so
\[
    \tn{Z}\geq\langle Z,f\rangle
    =\sum_{c\in\mathcal C}|\langle Z,f_c\rangle|.
\]
Taking expectations and using~\eqref{eq:sharp-pairing} gives
\begin{equation}\label{eq:sharp-centers}
    R\geq\frac{\sqrt{q'}}{8\sqrt6}\sum_{c\in\mathcal C}r_c.
\end{equation}

\emph{Recurrence and induction.}
Every site that is not good satisfies $r_i<8r_i(q')$. Combining this
with~\eqref{eq:sharp-packing} and~\eqref{eq:sharp-centers}, we obtain
\begin{equation}\label{eq:sharp-radii-recursion}
    L(q)\leq 8\sqrt6\,\frac{q}{\sqrt{q'}}R+8L(q').
\end{equation}
If $2\leq q<256$, then $q'=1$ and $q/\sqrt{q'}=q\leq16\sqrt q$.
If $q\geq256$, then $q'=\lfloor q/256\rfloor\geq q/512$, so
$q/\sqrt{q'}\leq\sqrt{512q}$. Set $C_0=512\sqrt{3}$. In both cases,
\begin{equation}\label{eq:sharp-radii-step}
    L(q)\leq (C_0/2)\sqrt q\,R+8L(q').
\end{equation}
For $2\leq q<256$, the last term is zero. For $q\geq256$, we have
$q'\leq q/256<q$. By the induction hypothesis,
$8L(q')\leq8C_0\sqrt{q'}\,R\leq(C_0/2)\sqrt q\,R$.
Strong induction now gives $L(q)\leq C_0\sqrt q\,R$ throughout the
stated range.
\end{proof}

\begin{proof}[Proof of Theorem~\ref{thm:sharp-stage}]
We first prove~\eqref{eq:sharp-bernoulli}. Fix $0<t\leq1$ and put
$q=\max\{2,\lceil 1/t\rceil\}$. Then $1/t\leq q\leq2/t$, so $tq\leq2$.
If $q\leq n/2$, Proposition~\ref{prop:recurrence} and
Lemma~\ref{lem:sharp-radii} give
\begin{align}
    H(t)&\leq 4tL(q)+24\sqrt{3/q}\,R
        +2\sum_{\ell\geq25}H(c_\ell t)\notag\\
    &\leq (4\sqrt2\,C_0+24\sqrt3)\sqrt t\,R
        +2\sum_{\ell\geq25}H(c_\ell t),\label{eq:sharp-sampling-step}
\end{align}
where $c_\ell=(4e/\ell)^\ell$. The second inequality uses
$t\sqrt q\leq\sqrt{2t}$ and $q^{-1/2}\leq\sqrt t$.

If $q>n/2$, put $q_*=\lfloor n/2\rfloor+1$. For any two sites $i,j$,
the closed balls about them with radii $r_i(q_*)$ and $r_j(q_*)$ each
contain more than half the sites. They therefore intersect, so
$d(i,j)\leq r_i(q_*)+r_j(q_*)$. Since $\tn{v_i}$ is the average
distance from $i$ to all sites, we have
\[
    \sum_i\tn{v_i}=\frac1n\sum_{i,j}d(i,j)
    \leq 2L(q_*)\leq 2C_0\sqrt{q_*}\,R\leq2C_0\sqrt n\,R.
\]
As $n/2<q\leq2/t$, we have $nt<4$. The triangle inequality gives
\[
    H(t)\leq t\sum_i\tn{v_i}
    \leq2C_0t\sqrt n\,R\leq4C_0\sqrt t\,R.
\]

Fix the server placement and define
$M_{1/2}=\sup_{0<u\leq1} H(u)/(\sqrt u\,R)$.
Since the sites are distinct and $n\geq2$, we have $R>0$ by
\eqref{eq:bench}. The bound $H(u)\leq nuR$ shows that $M_{1/2}\leq n$.
Since $c_\ell\leq2^{-\ell}$ for $\ell\geq25$,
\[
    2\sum_{\ell\geq25}\sqrt{c_\ell}
    \leq\frac{2\cdot2^{-25/2}}{1-2^{-1/2}}<\frac1{16}.
\]
The tail in~\eqref{eq:sharp-sampling-step} is therefore at most
$M_{1/2}\sqrt t\,R/16$. Since
$4C_0\leq4\sqrt2\,C_0+24\sqrt3$, the bound for $q>n/2$ also suffices.
Dividing by $\sqrt t\,R$ and taking the supremum gives
$M_{1/2}\leq4\sqrt2\,C_0+24\sqrt3+M_{1/2}/16$. Hence
\begin{equation}\label{eq:sharp-sampling-bound}
    H(t)\leq C_1\sqrt t\,R,\qquad
    C_1=\frac{16}{15}(4\sqrt2\,C_0+24\sqrt3).
\end{equation}

We next pass from independent sampling to subsets of fixed size.
For $1\leq j\leq k\leq n$, choose a uniform $k$-subset $F_k$ of $[n]$
and then a uniform $j$-subset $F_j$ of $F_k$. Each label in $F_k$
belongs to $F_j$ with conditional probability $j/k$, so
$\E[\nu_{F_j}\mid F_k]=\nu_{F_k}$. By convexity of the transport norm,
\[
    W_1(\nu_{F_k},\nu)\leq\E[W_1(\nu_{F_j},\nu)\mid F_k].
\]
Since the marginal law of $F_j$ is uniform over all $j$-subsets,
taking expectations gives $g_k\leq g_j$.

Fix $1\leq k\leq n$ and set $t=k/(2n)$. Write $K=K(t)$ for the sample
size defined in Subsection~\ref{sec:sampling}. Then
$K\sim\operatorname{Bin}(n,t)$ and $\E[K]=nt=k/2$. Since $h_j=jg_j$
for $1\leq j\leq n$ and $h_0=0$, the monotonicity just proved gives
\begin{equation}\label{eq:sharp-truncation}
    H(t)=\E[h_K]\geq g_k\,\E[K\mathbf{1}_{\{1\leq K\leq k\}}].
\end{equation}
Given $J_i(t)=1$, the event $K>k$ occurs precisely when at least $k$
of the other $n-1$ labels are sampled. Summing over $i$ and using
Markov's inequality, we obtain
\[
    \E[K\mathbf{1}_{\{K>k\}}]
    =nt\Pr\{\operatorname{Bin}(n-1,t)\geq k\}
    \leq nt\frac{(n-1)t}{k}\leq\frac{nt}{2}.
\]
Thus $\E[K\mathbf{1}_{\{1\leq K\leq k\}}]\geq nt/2=k/4$.
Substituting this in~\eqref{eq:sharp-truncation} and using
\eqref{eq:sharp-sampling-bound} gives
\begin{equation}\label{eq:sharp-fixed-size}
    h_k=kg_k\leq4H(k/(2n))\leq2\sqrt2\,C_1\sqrt{k/n}\,R.
\end{equation}

Finally, since $\sum_i v_i=0$, a subset and its complement give opposite
signed measures of the same norm. The complement of a uniform
$k$-subset is a uniform $(n-k)$-subset, so $h_k=h_{n-k}$, with
$h_n=h_0=0$. Gupta et al.~\cite[Appendix~C.3]{GGPW} use this symmetry,
and Harada~\cite[Lemma~4.1]{Harada} states the identity in terms of $g_k$.
Applying~\eqref{eq:sharp-fixed-size} at both sizes gives
$h_k\leq2\sqrt2\,C_1\sqrt{\min\{k,n-k\}/n}\,R$ for $1\leq k<n$.
Since $\min\{k,n-k\}\leq2k(n-k)/n$, we obtain
\[
    g_k=\frac{h_k}{k}
    \leq\frac{4C_1}{n}\sqrt{\frac{n-k}{k}}\,R
    \leq\frac{8C_1}{n}\sqrt{\frac{n-k}{k}}\,\E[\OPT_\nu],
\]
where the last inequality follows from~\eqref{eq:bench}. This proves the
claim for distinct sites with $C=8C_1$. For $k=n$, both sides are zero.

If several server labels share a location, use the perturbed metric
$d_\epsilon$ on $[n]$ from Subsection~\ref{sec:competitive}. For
$\epsilon>0$, let $g_k^\epsilon$ and $\OPT_\nu^\epsilon$ denote the
corresponding expected cost of a single match and the offline optimum
under $d_\epsilon$, respectively. As shown in the proof of
Theorem~\ref{thm:canonical},
\[
    0\leq g_k^\epsilon-g_k\leq\epsilon,\qquad
    0\leq\OPT_\nu^\epsilon-\OPT_\nu\leq n\epsilon.
\]
Applying the bound for distinct sites and letting $\epsilon$ tend to
zero proves the theorem for coincident server locations as well.
If $n=1$, the unique match costs zero.
\end{proof}

\paragraph{Sharpness of the exponent.}
The exponent $1/2$ of $k/n$ in~\eqref{eq:sharp-fixed-size} is sharp,
even when the bound is expressed in terms of $\E[\OPT_\nu]$.
Let $n$ be even and place $n/2$ servers at each of two sites at distance
one. Whichever site holds the last free server, half the request mass
must travel distance one, so $h_1=g_1=1/2$. Let
$Y\sim\operatorname{Bin}(n,1/2)$ count the requests at the first site.
Exactly $|Y-n/2|$ requests must cross between the two sites in an
optimal matching. Thus
\[
    \E[\OPT_\nu]=\E[|Y-n/2|]\leq\sqrt{\operatorname{Var}(Y)}
    =\frac{\sqrt n}{2}.
\]
If there were constants $C'>0$ and $\beta>1/2$ such that
$h_k\leq C'(k/n)^\beta\E[\OPT_\nu]$ for every instance and every
$1\leq k\leq n$, then at $k=1$ we would have
$1/2\leq (C'/2)n^{1/2-\beta}$. This fails as $n$ tends to infinity
through the even integers.

\end{document}